\RequirePackage{fixltx2e}
\documentclass[article]{revtex4-1}
\usepackage{comment}
\usepackage{subfig}
\usepackage{amssymb}
\usepackage{caption}
\usepackage[pdftex]{graphicx} 

\newcommand{\bea}{\begin{eqnarray}}
\newcommand{\eea}{\end{eqnarray}}

\begin{document}

\title{Vibrational Heat Transport in Molecular Junctions}

\author{Dvira Segal}
\affiliation{Chemical Physics Theory Group, Department of Chemistry, University of Toronto,
80 St. George Street Toronto, Ontario, Canada M5S 3H6}
\author{Bijay Kumar Agarwalla}
\affiliation{Chemical Physics Theory Group, Department of Chemistry, University of Toronto,
80 St. George Street Toronto, Ontario, Canada M5S 3H6}

\begin{abstract}
We review studies of vibrational energy transfer in a molecular junction geometry, consisting
of a molecule bridging two heat reservoirs, solids or large chemical compounds. This setup is of interest
for applications in
molecular electronics, thermoelectrics, and nanophononics,
and for addressing basic questions in the theory of classical and quantum transport.
Calculations show that system size, disorder, structure, dimensionality,
internal anharmonicities,
contact interaction, and quantum coherent effects, are factors that interplay
to determine the predominant mechanism (ballistic/diffusive),
effectiveness (poor/good) and functionality (linear/nonlinear) of thermal conduction at the nanoscale.
We review recent experiments and relevant calculations of quantum heat transfer in molecular junctions.
We recount the Landauer approach, appropriate for the study of elastic (harmonic) phononic transport,
and outline techniques which incorporate molecular anharmonicities.
Theoretical methods are described along with examples illustrating the challenge
of reaching control over vibrational heat conduction in molecules.
\end{abstract}

\maketitle

\tableofcontents

\section{INTRODUCTION}
\label{intro}

Achieving control over energy transfer at the molecular scale is essential for the realization
of molecular-based technologies.
Good thermal conductors are desired in electronic applications where it is necessary
to efficiently remove excess heat generated by electronic heat dissipation \cite{Pop,Shi,balandin11}.
On the other hand, materials which are poor conductors of heat are required
for the design of efficient thermoelectric devices \cite{Dresselhaus,Shakouri}.
Intramolecular vibrational energy redistribution (IVR) and
intermolecular vibrational energy transfer are central processes in chemistry,
fundamental to our understanding of chemical dynamics,  specifically, reaction rates \cite{VER,grueble04}.
Moreover, in biomolecules, energy transfer processes determine stability, function, and regulation  \cite{Leitner}.

\begin{figure}
  \begin{minipage}[a]{0.4\textwidth}
   \vspace{-40mm}
(a)\includegraphics[width=\textwidth]{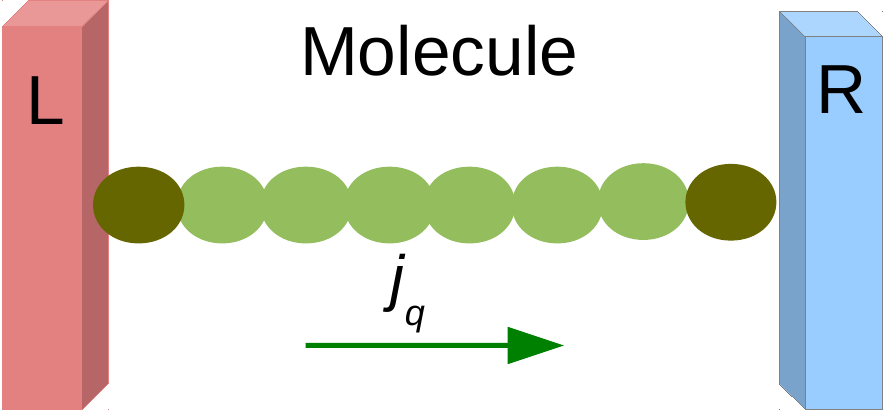}
  \end{minipage}
\hfill
  \begin{minipage}[b]{0.45\textwidth}
\vspace{5mm}  
(b)\includegraphics[width=\textwidth]{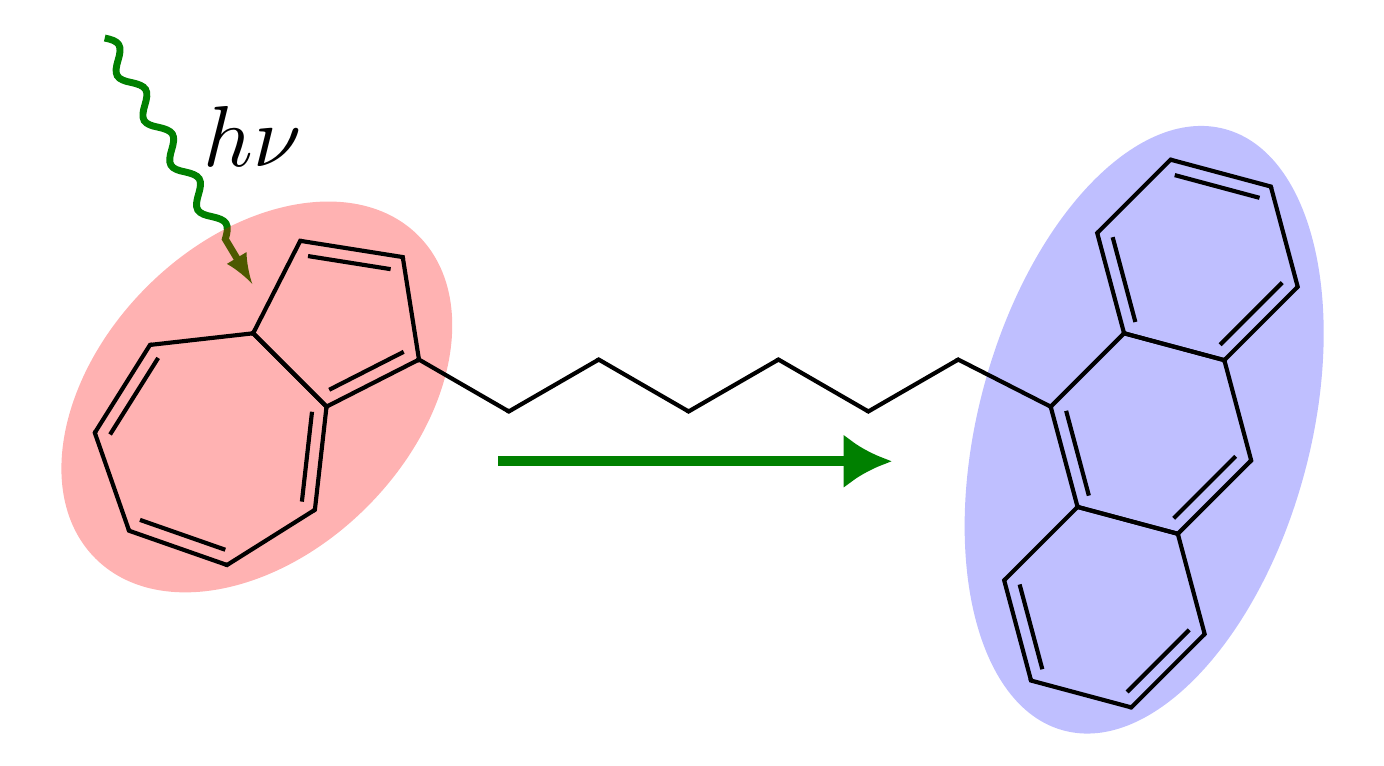}
  \end{minipage}
\caption{(a) Scheme of a molecular junction consisting of two solids
(possibly made of different materials, gold and quartz)
and a molecule of several units, typically functionalized at its ends to improve chemical bonding with the solids, e.g.,
alkane-thiols of $N$ units.
Heat current is flowing from the hot reservoir (e.g., the left one) to the other-cold side.
(b) Vibrational energy flow in molecules can be measured 
using pump-probe laser spectroscopy tools, in liquid solutions at ambient temperatures.
In this illustration, the molecule of interest (alkane chain) is connected at its two ends to
large molecular groups acting as local heat reservoirs, for example, azulene and anthracene \cite{Troe04}.
The azulene group is excited optically. The time
for intramolecular vibrational energy transfer, through the bridge to the anthracene unit, is measured for different chain lengths,
to identify the dominant transport mechanism.
}
\label{figModel}
\end{figure}

In this review, motivated by  progress in probing heat transfer at the nanoscale,
particularly, in self-assembled monolayers  (SAMs) and nanostructured materials \cite{cahill13},
we focus on the problem of vibrational heat flow through molecular junctions.
After the presentation of representative experimental results,
we describe quantum mechanical-based methodologies for the simulation of vibrational energy flow,
focusing on the behavior of a single molecule.
In the generic setup of interest a molecule is placed
between two macroscopic contacts, labeled by $L$ and $R$, see Fig. \ref{figModel}(a) for a schematic illustration.
These thermal baths are characterized by well-defined thermodynamic properties, essentially, their temperatures $T_L$ and $T_R$,
potentially creating large biases, $(T_L-T_R)/(T_L+T_R)\sim1$.
Considering transport of vibrational energy in the junction, basic quantities of interest are the heat current $j_q$
and the thermal conductance $\kappa\equiv j_q/\Delta T$ (units W/K) with $\Delta T=T_L-T_R$
as the temperature difference between the bulk objects.
Complementary to solid/molecule/solid junctions, heat transfer in
molecular chains can by studied in a ``bridged" configuration,
by connecting a molecule at its two ends to relatively large side groups which act (approximately)
as thermal baths, see Fig. \ref{figModel}(b) for a scheme following Ref. \cite{Troe04}.
In such experiments, a short laser pulse is applied to heat one end of the extended molecule.
The heat transfer rate is then identified from optical-spectroscopy means, by measuring either the rate of energy loss from the
excited unit, or energy gain at the other end.

Besides its relevance to technologies such as molecular electronics
\cite{Ratnerforty,Latha13}, phononics \cite{phononic,Li-AIP}, thermoelectrics \cite{Shakouri},  spin caloritronics \cite{spinC},
Nanoelectromechanics \cite{mems}, quantum optomechanics \cite{optoM} and quantum information processing 
with hybrid quantum circuits  \cite{Nori},
the generic model displayed in Fig. \ref{figModel}(a) is of basic interest for
probing open quantum systems phenomena.
Energy transfer processes in molecular junctions involve many-body interactions and  quantum effects in a
non-equilibrium situation, thus the setup
can be employed to address fundamental questions in statistical mechanics, condensed matter physics,
thermodynamics, and classical and quantum mechanics.
Is energy transport in a given system ballistic or diffusive?
What are the roles of many-body interactions in determining transport characteristics,
especially when the system is driven beyond linear response?
What are the signatures of quantum effects in molecular-scale heat transport?
Can we design or predict a certain functionality from the model Hamiltonian?
Other questions (not reviewed here), which have
attracted significant attention, concern the emergence of the phenomenological-macroscopic Fourier's law of
heat conduction from first principles (classical and quantum) \cite{dhar08},
and the derivation of sufficient or necessary conditions for realizing negative differential thermal
conductance and thermal rectification \cite{phononic,Li-AIP,casati-diode}.
The latter effect, also referred to as the thermal diode effect, features junctions in which
the magnitude of the thermal current depends on the direction (polarity) of the temperature difference.

In this review our main focus are theoretical frameworks for the calculation
of vibrational energy flow in molecular junctions, and we contain ourselves to quantum mechanical treatments.
Why concern with quantum effects? Molecular vibrational frequencies typically extend
room temperature, $\hbar \omega>k_BT$, thus the quantum statistics is important;
in classical simulations the average energy per phonon mode is $k_BT$, with $k_B$ as the Boltzmann constant,
resulting in an overestimate of
e.g., the specific heat (thus the thermal conductivity), since $\frac{\hbar \omega}{e^{\hbar\omega/k_BT}-1}<k_BT$.

Because of the wide scope of the field of heat transfer, we leave out many interesting
topics which were covered in recent reviews.
Concepts, experiments, and computational approaches in nanoscale thermal transport
were recently organized into several comprehensive works \cite{cahill13,Li-AIP,GChen,GoodsonCNT}, with a focus on
low-dimensional nanostructured materials such as graphene, carbon nanotubes, and Si nanowires,  
rather than molecular-level transport.
Other relevant reviews cover 
control over energy dissipation at the nanoscale \cite{Pop},
advances in thermoelectrics of semiconductor nanostructured materials \cite{Shakouri},
studies over the nature of vibrational energy flow in proteins \cite{Leitner}, and
heating, heat transfer, and thermoelectricity at the nanoscale \cite{dubi}.
Concerning methodologies for calculating nanoscale phonon transport,
recent reviews detailed classical simulations \cite{dhar08,Lepri03,MDphonon} 
and quantum-mechanical methodologies \cite{dhar08},  specifically 
Green's function-based approaches  \cite{wang-lu08,wang14,mingo}.
Other studies featured phononic devices \cite{phononic, Li-AIP}
and phonon-assisted effects in molecular electronic conduction \cite{nitzan}. 
Obviously, the interaction of electrons with phonons is intrinsic
for both charge and energy transfer processes in molecular systems \cite{nitzaneph}. Here we simplify the problem
considerably by studying only the dynamics of vibrational-nuclear degrees of freedom.
Even under this separation, the situation depicted in Fig. \ref{figModel} is deeply complex given
the interplay of quantum effects, many-body (phonon-phonon) interactions, far-from-equilibrium driving, and the
diversity in molecular structures and contact geometries. 

Before discussing experimental results and theoretical methodologies, it is useful to recall
two central-limiting energy transfer mechanisms which are well defined in macro-to-mesoscale systems: ballistic and diffusive.  
Ballistic transport corresponds to  
direct point-to-point propagation of energy, obeying the scaling 
$l\propto t$, with $t$ as the traveling time to cross the distance  $l$.
It shows up in ordered structures in which the mean free path,
the average distance traveled by phonons between successive scattering events,
is larger than the size of the conductor. This is the case e.g., in some (up to $\mu m$-length) nanotubes \cite{balandin11},
and more relevant to our review,
alkane chains of 5-25 units, at room temperature
\cite{Troe04,Wang06,Gotsmann14, Cahill12, DlottRev,rubtsov15,rubtsovRev15}.
Diffusive energy transport is a multiple-step process, with energy ``hopping" between sites, obeying the scaling 
$l\propto \sqrt{t}$.
It is observed in bulk systems and disordered structures, e.g., peptides \cite{hammpnas07,Leitner}, 
glasses, amorphous and doped nanostructures \cite{cahill13}. 

The distinction between ballistic and diffusive dynamics, and the notion of a mean free path, are natural to
macro-scale and nanostructured systems, but in relatively short molecules as considered below (5-50 \AA) 
these concepts are not well defined.
Instead, in the molecular realm we distinguish between harmonic models,
assuming atomic displacements follow Hooke's law, and anharmonic cases, allowing for anharmonicities in the molecular force field.
In harmonic models the current  may show ballistic characteristics, with the conductance
determined by the contacts only, or, ``phonon tunneling", when phonons off-resonance with molecular vibrations,
cross the junction, showing features of quantum tunneling \cite{segal03}.
We use Eq. (\ref{eq:jqsingle}) below to exemplify the concepts of ballistic and tunneling phonon dynamics in molecular 
heat conduction. 
Note that we use the notions of phonons and vibrations interchangeably, 
referring to the collective motion, excitations, of atoms.


\section{EXPERIMENTS: FOCUS ON HEAT CONDUCTION IN ALKANE CHAINS}

We begin with the presentation of several experiments, corresponding to the schemes in
panels (a) and (b) of Fig. \ref{figModel}. This introduction
serves to motivate modeling and computational works, but it does not aim to give readers a complete review
over heat transfer measurements, provided e.g. in Refs. \cite{cahill13,GChen}.
We focus on studies of length-dependent thermal transport in
alkane chains with 2-25 methylene units.
This molecule serves as a standard for phononic transport measurements for several reasons.
First, these sigma-bond hydrocarbons are poor conductors of electricity, therefore
thermal conduction takes place by phonons, not electrons.
Second, the molecule is uniform and quasi one-dimensional. Lastly,
at room temperature
anharmonic effects are expected to be non-influential even in long chains of $N\sim 100$ units \cite{segal03}.
Altogether, alkane chains with $N \gtrsim 5$ units are expected to conduct vibrational energy ballistically at room temperature, as confirmed by simulations \cite{segal03} and experiments
\cite{Troe04,Wang06,Gotsmann14, Cahill12, DlottRev,rubtsov15,rubtsovRev15}.

\begin{figure}[htbp]
\vspace{5mm}
  \begin{minipage}[a]{0.45\textwidth}
   \vspace{-60mm} 
(a)\includegraphics[width=\textwidth]{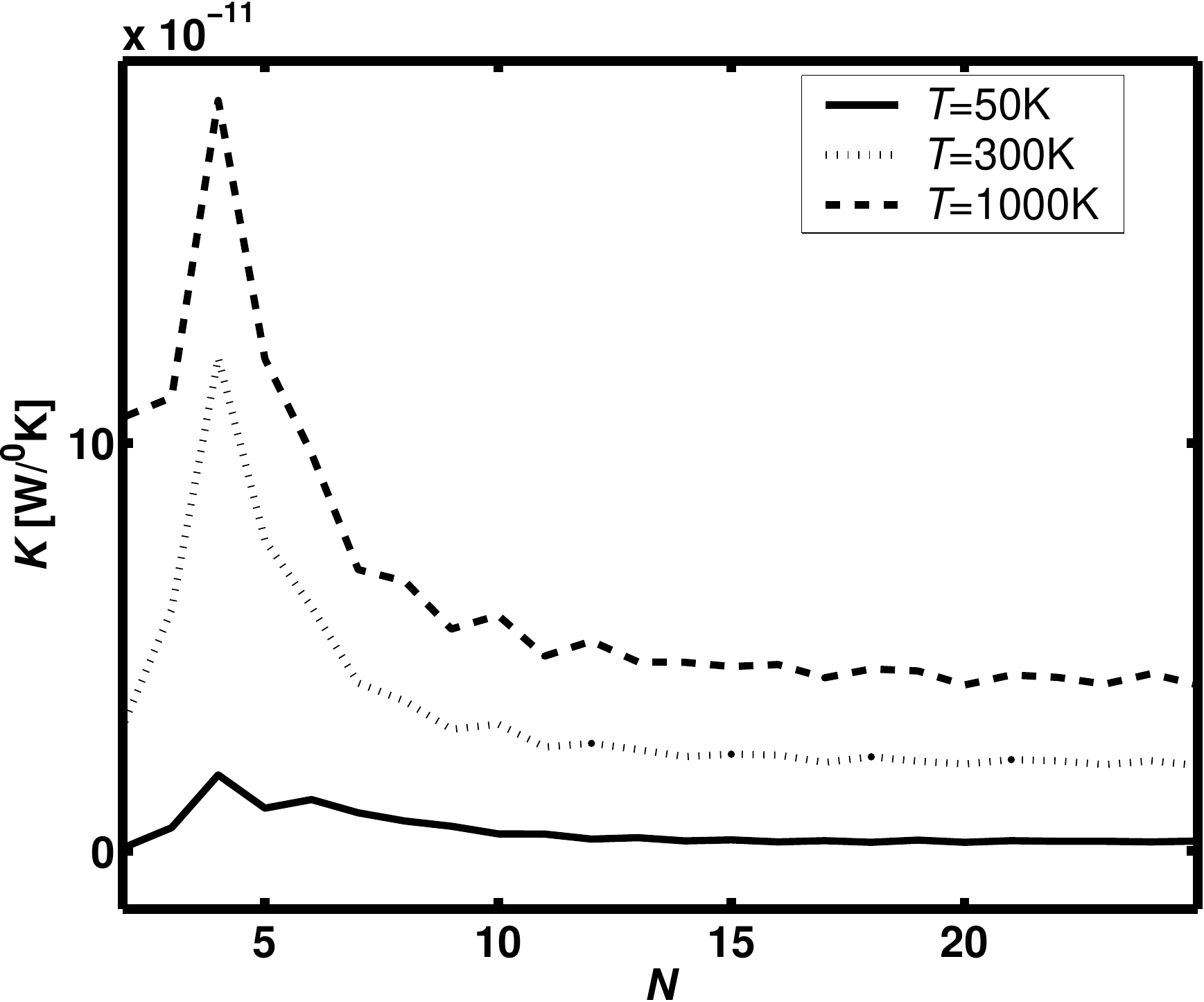}
  \end{minipage}
\hfill
  \begin{minipage}[b]{0.42\textwidth}
\hspace{1em} (b) \includegraphics[width=\textwidth]{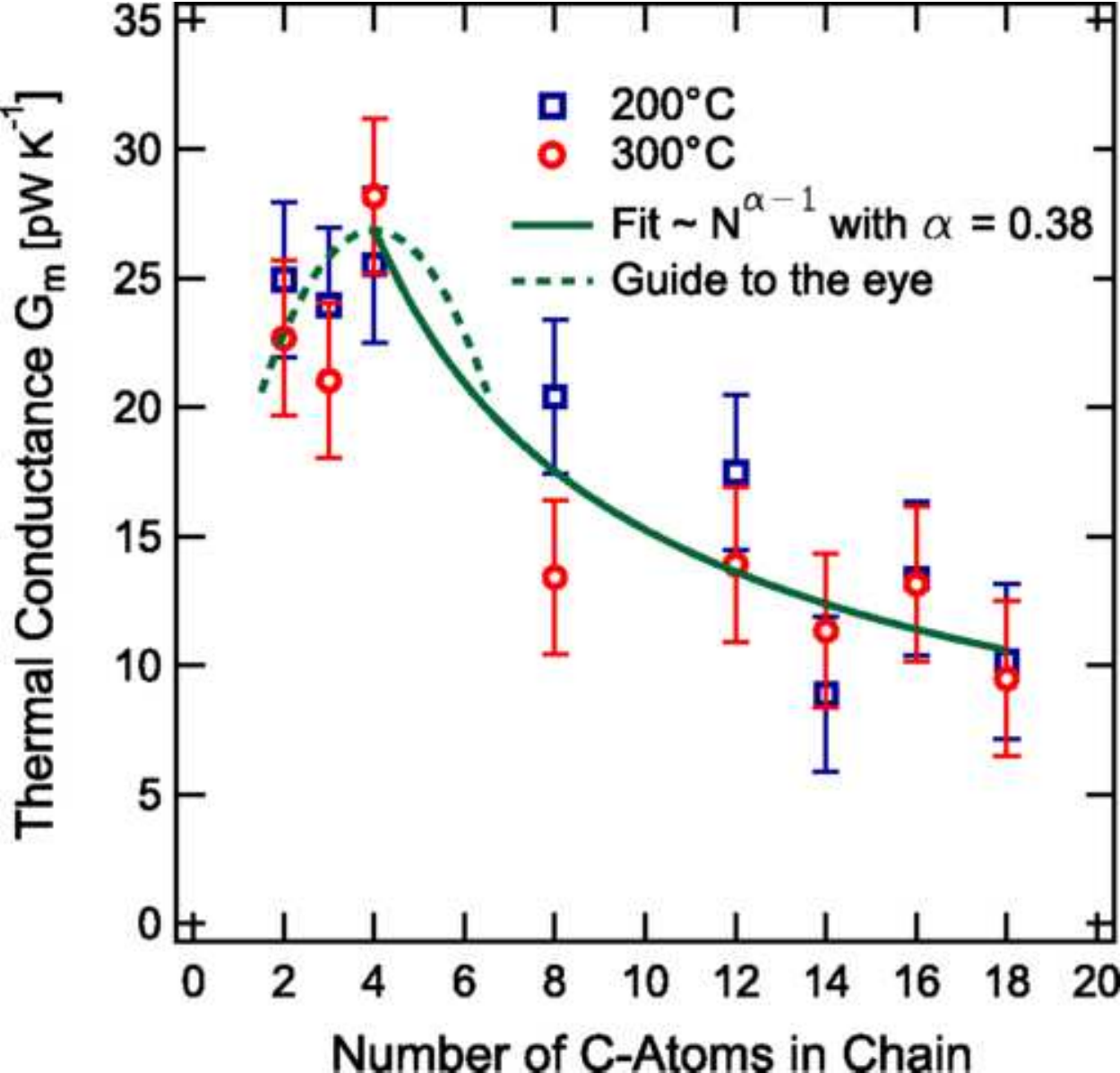}
  \end{minipage}
\vspace{3mm}
\caption{
Thermal conductance of alkane chains with increasing size.
(a) Simulations under the harmonic approximation based on the Landauer formula, as reported in Ref. \cite{segal03}. 
The reservoirs' spectral functions are modeled by $\gamma(\omega)=\frac{a}{\hbar}e^{-\omega/\omega_c}$,
identical for $L$ and $R$, with Debye (cutoff) frequency $\omega_c$= 0.05 eV 
and a coupling energy $a=$ 1 eV. 
Full line: $T$=50 K; dotted  line: $T=$ 300 K;  dashed line: $T$=1000 K.
(b) Measured values for the thermal conductance of SAMs of alkane thiols [HS-(CH$_2$)$_{N-1}$-CH$3$]
on gold substrates on Mica, using SThM with a silicon tip.
Panel (b) is reprinted with permission from Reference \cite{Gotsmann14}. 
Copyright 2014 American Physical Society.
}
\label{figAlkane}
\end{figure}

Solid-SAMs-solid junctions were studied in several works. 
The thermal conductance of
Au-alkanedithiol-GaAs SAMs with 8-10 carbon units was measured in Ref. \cite{Wang06}  with a $3\omega$ technique, resulting in
the room-temperature value (per unit area) of $\kappa \sim 25-28$ MW m$^{-2}$ K$^{-1}$,
independent of length.
More recently, heat transport measurements with scanning thermal microscopy (SThM)
were reported in Ref. \cite{Gotsmann14}, with an alkane-thiol molecule adsorbed on a gold surface
and a silicon tip approaching the SAM from above.
The measured value for the thermal conductance is depicted in Fig. \ref{figAlkane}(b).
It shows a non-monotonic behavior: The conductance first rises with length for short chains, with the maximum conductance of
25 pW/K for a chain with four carbon atoms, but in longer chains $\kappa$ is reduced to
$\sim 10$ pW/K.
This crossover behavior appears in simulations, see Fig. \ref{figAlkane}(a),  
to be discussed in more details in Sec. \ref{exampleH}.

In ballistic transport the conductance reflects the molecule-contact interaction
rather than intrinsic molecular properties.
The effect of chemical bonding on interfacial heat transport in Au/alkane SAMs/Quartz junctions was
examined in Ref. \cite{Cahill12}, by studying chains with either methyl or thiol-termination, using the time-domain
thermoreflectance (TDTR) technique, an optical pump-probe tool.
The thiol-ending group forms a stronger (covalent) bond to the Au surface,
relative to the weak (van der Waals) forces attaching a methyl group to a gold
substrate. In accord with expectations, 
the interface thermal conductance was enhanced with the increase of bond strength:
The interfacial thermal conductance (per unit area) measured for the thiol ending group was $\kappa$=68 MW m$^{-2}$ K$^{-1}$.
A lower value was reached for the interface missing thiol bonds,  $\kappa=36$ MW m$^{-2}$ K$^{-1}$ \cite{Cahill12}.
In Ref. \cite{malen} it was also demonstrated experimentally that 
the thermal conductance of alkane SAMs can be systematically controlled-reduced by using metals with 
mismatched phonon spectra, e.g., Au and Pt.

Bridged-type situations as depicted in Fig. \ref{figModel}(b) were examined in several works, 
revealing information on mechanisms of heat flow from the dynamics of the transients.
In one of the earliest studies, alkane chains were connected to large chromophores, 
azulene and anthracene, which effectively act as heat reservoirs \cite{Troe04}.
The IVR process began by the electronic excitation of the azulene end-group, and
after a fast internal conversion,
vibrational energy localized on the azulene propagated
through the bridge to the anthracene, with an IVR time constant of $\tau\sim 5-6$ ps.
It was demonstrated that results were consistent with a ballistic transport model \cite{Troe04}.
At later times, equilibration with the surrounding solvent took place.

A rather different approach, time-resolved flash-heating of SAMs on substrates, was utilized in Ref. \cite{dlottscience},
see also \cite{DlottRev,dlottCP}.
In this type of experiments, a thin metal film (gold) was heated by several hundreds of degrees within
a very short time ($\sim$ 1ps) using a laser pulse, to excite electrons near the metal surface.
Subsequently, the lattice temperature rose due to electron-phonon coupling, and heat propagated from the base of the SAM,
through the molecular chains, to the top (methyl) groups where it was detected
with vibrational sum-frequency generation (SFG) spectroscopy.
This signal is sensitive to disorder on the SAM, induced here by the temperature increase, as well as to
frequency shifts. By varying the length of the alkane chain it was determined that
heat flew in alkanes ballistically, with a velocity of $9.5$ \AA/ps,
to yield a thermal conductance (per molecule) of 50 pW/K. 

\begin{figure}
\includegraphics[scale=0.5]{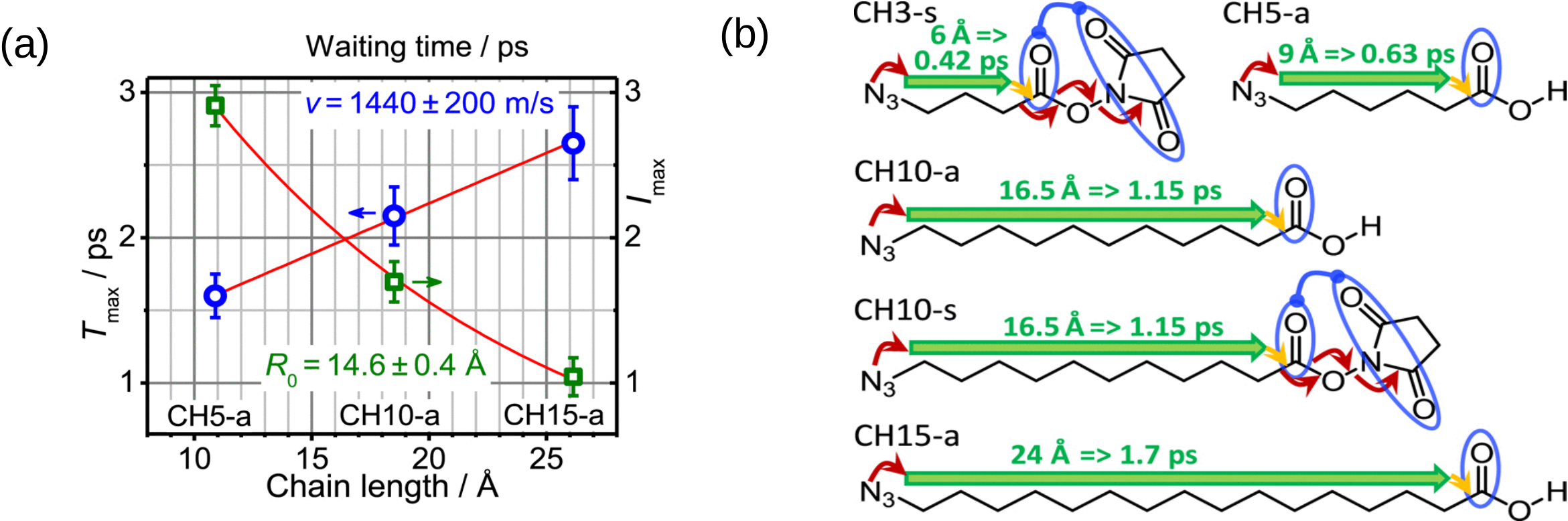}
\vspace{2mm}
\caption{Two-dimensional IR spectroscopy of vibrational energy trasnfer in alkanes.
(a) Inferred
energy transport time $T_{max}$, from the tag mode to the reporter,
and the tag-reporter cross-peak maximum amplitude $I_{max}$, plotted as a function of chain length.
The behavior of $T_{max}$ is approximated by a linear function, and from the (inverse) slope one receives the 
speed $\sim$ 14.40 \AA/ps.
The cross peak amplitude is approximated by an exponentially decaying function with a characteristic 
decay distance $R_0\sim 14.6$ \AA, representing the mean free path.
(b) Schemes of energy transport mechanisms in alkane chains with different reporter groups (blue).
Green arrows mark regions of ballistic propagation, red arrows mark (diffusive) steps from the tag group to the molecule,
and from the molecule to the reporter.
Yellow arrows point to the relaxation of the vibrational wavepacket.
Also indicated are ballistic transport times, computed from the chain length divided by the energy transport speed 1440 m/sec.
Reprinted with permission from Reference \cite{rubtsov15}.
Copyright (2015), AIP Publishing LLC. 
}
\label{figrubtsov}
\end{figure}

Other recent studies further considered a bridged setup in solution
and confirmed the ballistic energy transport mechanism at room temperature in alkane chains of 5-15 units, 
by using two-dimensional infrared spectroscopy (2DIR) \cite{rubtsov15,rubtsovRev15}.
In this experiment, the energy transfer process was initiated by vibrationally exciting
a ``tag" group attached to one end of the molecule (e.g. N$\equiv$N bond),
while the vibrational energy arriving at the other end was recorded by a ``reporter" group, e.g., a carbonyl.
Measurements at room temperature, in solution, revealed fast energy transport through alkanes with a
speed of 14.4 \AA/ps and a mean free path of 14.6\AA, see Fig. \ref{figrubtsov}.
Interestingly, while it was demonstrated that transport via the chain was ballistic,
it was argued that heat propagated through the end-groups in a diffusive manner \cite{rubtsov15}.
It was also demonstrated in Ref. \cite{RubtsovJPC} that by initiating dynamics with different tags,
different chain bands (e.g., CC stretching, CH$_2$ twisting modes) contribute to the transfer process.

Thermal transport experiments on hydrocarbons confirmed ballistic dynamics,
the predominant mechanism in short-ordered systems.
In contrast, diffusive dynamics is prevalent in complex systems (even of reduced dimensions),
nanotubes and nanowires \cite{balandin11,GoodsonCNT,Kim01}, polymers \cite{cahill13}, 
peptides \cite{hammpnas07,Leitner}, as well as in small organic molecules \cite{rubtsovRev15}.
For example, in Ref. \cite{hammpnas07} diffusive transport of vibrational energy was
observed in short 3$_{10}$-helical peptides. The process was
initiated by depositing excess energy
e.g. onto a vibrational CD mode. Flow of energy through
the helix was then detected optically by planting
vibrational probes, isotopically-labeled CO modes, in several locations away from the source, serving
as local thermometers. Measurements in Ref. \cite{hammpnas07} were consistent with a diffusion model,
with a diffusivity constant of 2 \AA$^2$/ps \cite{hamm12}.

In the following discussion over theoretical-computational frameworks
we do not direct the issue of the ballistic-diffusive crossover in thermal conduction, as
diffusive dynamics is not expected to fully develop
in short molecular junctions, our focus here.
More generally though we will evaluate
the role of nonlinearities, deviations from the harmonic approximation for atomic displacements, in the heat transport behavior.
What are the signatures of anharmonicities in relatively small phononic conductors?
First, one may observe the suppression of the thermal conductance with temperature, at high temperatures, 
when inelastic phonon scattering effects are at play \cite{juzarQME}. 
Further, anharmonicities can furnish nonlinear functionalities such as 
the thermal diode effect \cite{phononic,segal05SB,wuPRL09},
which is specifically predicted to develop in the azulene-alkane-anthracene compound of Fig. \ref{figModel}(b)
at high temperatures \cite{leitner-interface}.
Other distinct signatures of vibrational anharmonicities are processes such as energy localization and unidirectional
energy flow in molecules.  
Using IR Raman spectroscopy, it was manifested that in nitrobenzene 
vibrational energy could leave benzene-localized modes  to occupy global modes,
but energy could not propagate away after a nitro excitation \cite{dlottdiode1,dlottdiode2,dlottdiode3}.
Asymmetries in forward-backward transfer efficiencies certainly break the noninteracting phonon picture.
We describe now quantum mechanical approaches for modelling vibrational energy flow in molecular junctions.

\section{METHODOLOGIES: Harmonic Approximation and the Landauer Formula}
\label{theory-harmonic}

\subsection{Formalism}
\label{formalismH}

Under the harmonic approximation for atomic displacements,
the molecule (``subsystem") includes harmonically-connected masses, and it
is linked linearly-harmonically to thermal environments
composed  of harmonic oscillators (``baths"). 
Since all terms are harmonic, one can diagonalize the model Hamiltonian
to reach its fundamental normal modes (phonons) and solve the dynamics of this noninteracting problem exactly.
The steady state transport behavior of the model can be arrived at by using different methods,
including the generalized Langevin equation \cite{dhar08,segal03,talkner,roy06,kirkQLE} or, equivalently, the
non-equilibrium Green's function (NEGF) approach \cite{wang-lu08,mingo,ciraci01} (derivations under
the weak-coupling approximation can be found e.g., in Ref. \cite{geller}.)
Both methods yield a Landauer-type formula
for the heat current, an analogue of the
elastic-coherent expression for charge current \cite{landauer}.

Following Refs. \cite{dhar08,segal03, roy06}, we outline the derivation of the Landauer heat transport
expression from the principles of the
Langevin equation approach. In this procedure, one begins with Heisenberg equations of motion for
displacement and momentum operators,
assuming a system-bath factorized initial condition at $t=-\infty$ with the
reservoirs prepared in a thermodynamic equilibrium.
It can be shown that after integrating out the baths' coordinates,
the subsystem's displacements follow a generalized (quantum) Langevin equation
in which the effect of the reservoirs is encapsulated
within dissipation kernels and noise terms.
These terms particularly depend on
the baths' spectral properties.
Fourier transforming the generalized Langevin equation administers the steady state limit.
In frequency domain, the linear equations for molecular coordinates
can be readily solved to construct two-time correlation functions, to organize 
 a closed-form expression for the heat current, a Landauer-type formula
\cite{dhar08,segal03,roy06}.
For harmonic models, the quantum Langevin equation and the
NEGF method can be further employed to yield
closed expressions for the fluctuations of the heat current \cite{keiji07,bijayFCS12}.

We now provide the working expressions. The Hamiltonian for a harmonic junction reads
\bea
H=H_S+ H_L+H_R + H_I,
\label{eq:H1}
\eea
where the molecule $H_S$, the two contacts $H_{\nu}$, $\nu=L,R$, and the coupling term $H_I$ are written as
\bea
H_{\alpha} &=& \frac{1}{2} p_{\alpha}^T \, p_{\alpha} +  \frac{1}{2} u_{\alpha}^T\, K_{\alpha} \, u_{\alpha}, \quad \alpha=S, L, R  \nonumber \\
H_{I} &=& \sum_{\nu=L,R} u_{\nu}^T \,\Lambda_{\nu S}\, u_{S}.
\label{eq:HH}
\eea
The vector $u_{\alpha}= \sqrt{m_{\alpha}} x_{\alpha}$ gathers mass-normalized displacement operators
for the three regions $\alpha=L,R,S$. Similarly, $p_{\alpha}$ holds the conjugate momenta,
$T$ stands for matrix transpose.
The force constants are organized into real symmetric matrices $K_S$, $K_L$ and $K_R$, and
the harmonic (separable) interaction of the molecule with the two reservoirs is given in terms of
the force constant matrices $\Lambda_{LS}=[\Lambda_{SL}]^T$ and $\Lambda_{RS}=[\Lambda_{SR}]^T$.
Note that in general one does not need to assume a normal mode representation for the reservoirs as
this may be achieved through diagonalization with unitary matrices $U_\nu$,
\bea
U_\nu^{T}\, K_{\nu} \, U_\nu= \Omega_\nu^2. 
\eea
In  steady state, the phonon current across harmonic junctions
is given by a Landauer-type expression \cite{kirczenow}
\bea
j_q= \frac{1}{2\pi} \int_0^{\infty} d\omega \, \hbar \omega\, \mathcal T(\omega) \,[n_L(\omega)-n_R(\omega)].
\label{eq:land}
\eea
Here, $n_\nu(\omega)=[e^{\beta_\nu \hbar \omega}-1]^{-1}$ are Bose-Einstein distribution function for
the $\nu=L,R$ bath with the inverse temperature $\beta_\nu=1/(k_B T_\nu)$,
$\mathcal T(\omega)$ stands for the transmission function calculated from \cite{dhar08,Caroli}
\bea
{\cal T}(\omega)= {\rm Tr} \big[G_{0}^{r}(\omega) \Gamma_L(\omega) G_{0}^{a}(\omega) \Gamma_{R}(\omega)\Big].
\label{eq:Thar}
\eea
For small temperature differences, $\Delta T << T$ with $\Delta T\equiv T_L-T_R $, $T=(T_L+T_R)/2$
as the average temperature, the Landauer expression reduces to
\bea
j_q=\frac{\Delta T}{2\pi}\int_0^{\infty} d\omega \, \hbar \omega  \,\mathcal T(\omega) \,\frac{dn(\omega)}{dT},
\label{eq:landkappa}
\eea
with $n(\omega)$ as the Bose-Einstein function at the averaged temperature $T$.
The ingredients in Eq. (\ref{eq:Thar}) are the Green's function of the molecule,
$ G_{0}^{r}(\omega)= (G_{0}^{a}(\omega))^{\dagger}$, and its self energy,
\bea
G_0^r(\omega) &=& \big[(\omega+i0^{+})^2 -K_S -\Sigma_L^r(\omega)-\Sigma_R^r(\omega)\big]^{-1},
\nonumber \\
\Sigma_{\nu}^r(\omega) &=& [\Lambda_{\nu S}]^T g_{\nu}^r(\omega) \Lambda_{\nu S}, \,\,\,\,\,\,\,
\Gamma_{\nu}(\omega)= i \big[\Sigma_{\nu}^r(\omega)-\Sigma_{\nu}^a(\omega)], \quad \nu=L,R
\eea
where $\Gamma_{\nu}(\omega)$ is referred to as the spectral function for the leads.
The unperturbed-isolated  Green's function for the baths are given by
\bea
g_\nu^r(t)= -\theta(t) \, U_\nu \frac{\sin (\Omega_\nu t)}{\Omega_\nu} U_\nu^{T} , \,\,\,\,\,\,
g_\nu^r(\omega)=\int_{-\infty}^{\infty} e^{i\omega t} g_\nu^r(t) dt
= [(\omega+i0^+)^2-K_{\nu}]^{-1},
\eea
with $\theta(t)$ as the Heaviside step function.
The Landauer equation (\ref{eq:land}) relies on the harmonic approximation,
yet even under this strong assumption it can be employed to address a broad range of problems:
the behavior of heat conduction in one-dimensional chains, as opposed to the three-dimensional case,
the role of disorder in thermal transport, as well as impurities, molecular structure (linear or T-shaped objects), reservoir's spectral properties, and contact geometry.


We now specify  Eq. (\ref{eq:HH}) and introduce a chain model
with the first (last) atoms in the chain coupled to the left (right) baths,
\bea
H&=& H_S 
+
\sum_{l\in L} \left[ \frac{p_l^2}{2} + \frac{1}{2}\omega_l^2\left(u_l-\frac{\lambda_lu_1}{\omega_l^2}\right)^2
\right]
+\sum_{r\in R} \left[ \frac{p_r^2}{2} + \frac{1}{2}\omega_r^2
\left(u_r-\frac{\lambda_ru_N}{\omega_r^2}\right)^2 \right].
\label{eq:H2}
\eea
%
%
The small-letter parameters,  $\lambda$, $\omega_{l,r}$, are elements in the matrices  $\Lambda$, and $\Omega_{\nu}$, 
respectively. 
The self energy matrices reduce to single elements, e.g., at the left contact
\bea
[\Sigma_L^r(\omega)]_{n,n'} =
\Sigma_L^r(\omega) \delta_{n,n'}\delta_{n,1}=
- \int_0^{\infty}e^{i\omega t}
\sum_{l \in L} \lambda_l^2 \frac{\sin (\omega_l t)}{\omega_l}dt.
\eea
The real part of the self energy presents frequency shifts to molecular modes,
the result of the coupling to the bath. 
The imaginary part of the self energy is responsible for
energy damping. Ignoring frequency shifts, it is convenient to express it
in terms of a (real-valued) function $\gamma_{\nu}(\omega)$, defined from
$\Sigma_\nu^r(\omega)\equiv-i\omega \gamma_{\nu}(\omega)$. It corresponds to the friction
coefficient in the language of the quantum Langevin equation.

Simulations of phononic heat current based on Eq. (\ref{eq:land}) 
are feasible, but analytic results are limited
to special cases, e.g, to the classical-high temperature limit of a uniform chain \cite{dhar08},
or to short atomic bridges \cite{kirk14}.
To provide insights into the transport behavior, we exemplify Eq. (\ref{eq:land}) in the single-atom limit,
taking into account a single oscillator which is
directly coupled to the $L$ and $R$ reservoirs.
We denote the frequency of this  mode by $\omega_0$
adhering to the notation employed in the literature, see e.g. Refs. \cite{SegalME06,Segalradiation}.
%
The transmission function of this bridge is given by
\bea
\mathcal T(\omega) =
\frac{4 \omega^2\gamma_L(\omega)\gamma_R(\omega)}{(\omega^2-\omega_0^2)^2 + \omega^2[\gamma_L(\omega)+\gamma_R(\omega)]^2 },
\label{eq:Tw0}
\eea
and the phonon current through the junction is
\bea
j_q=\frac{2}{\pi} \int_0^{\infty} d\omega \hbar \omega \, \frac{\omega^2\gamma_L(\omega)\gamma_R(\omega)}{(\omega^2-\omega_0^2)^2 + \omega^2[\gamma_L(\omega)+\gamma_R(\omega)]^2 } [n_L(\omega)-n_R(\omega)],
\label{eq:jqsingle}
\eea
see also Refs. \cite{kirk14,schiller} (albeit adopting a different notation).
This simple expression describing quantum coherent transport allows us to clarify several concepts: 
%
(i) When incoming phonons are in resonance with the molecular mode, $\omega=\omega_0$, the transmission is only limited by contact effects.
We refer to this type of motion as ``ballistic" transport.
Furthermore, in the ideal case with a perfect transmission, ${\mathcal T}(\omega)=1$, 
we evaluate the conductance from (\ref{eq:landkappa}) and receive 
$\kappa=\frac{1}{2\pi}\int_0^{\infty} d\omega \hbar\omega \frac{dn}{dT}=k_B^2 \pi^2 T/3h$,
which is the (universal) quantum of thermal conductance
\cite{ciraci01,kirczenow,pendry,Wulf}, confirmed experimentally in Ref. \cite{kappaQ}.
(ii) The damping function $\gamma_{\nu}$, resulting from the molecule-bulk coupling, broadens the resonance allowing for
``tunneling" of phonons, transmission of modes off-resonance with the molecular frequency, $\omega\neq\omega_0$.
It can be shown that the tunneling contribution exponentially decays
with bridge-length \cite{segal03}, in analogy with the super-exchange mechanism of electron transfer \cite{super}.
(iii) 
The conductance grows with $T$ at low temperatures,
but once $k_BT>\hbar \omega_0$, it saturates as it approaches the classical limit.

\begin{figure}
  \begin{minipage}[a]{0.45\textwidth}
   \vspace{-90mm}
\includegraphics[width=\textwidth]{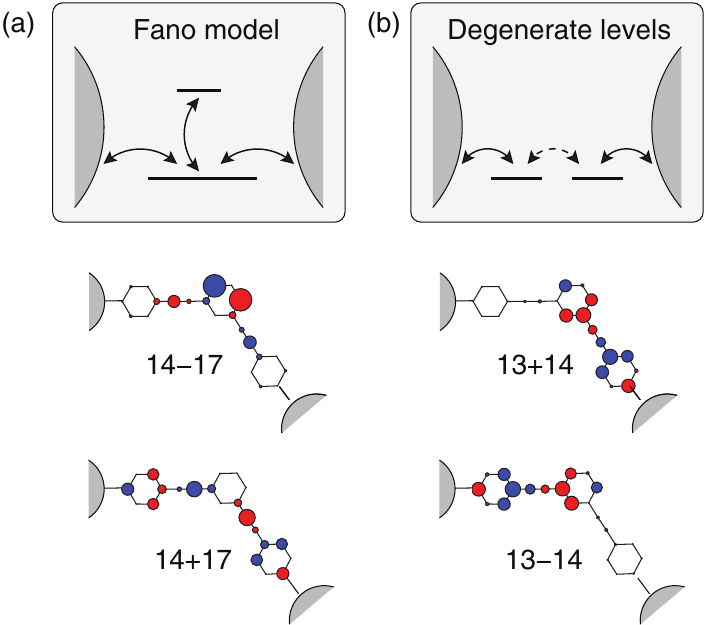}
  \end{minipage}
\hfill
  \begin{minipage}[b]{0.5\textwidth}
\includegraphics[width=\textwidth]{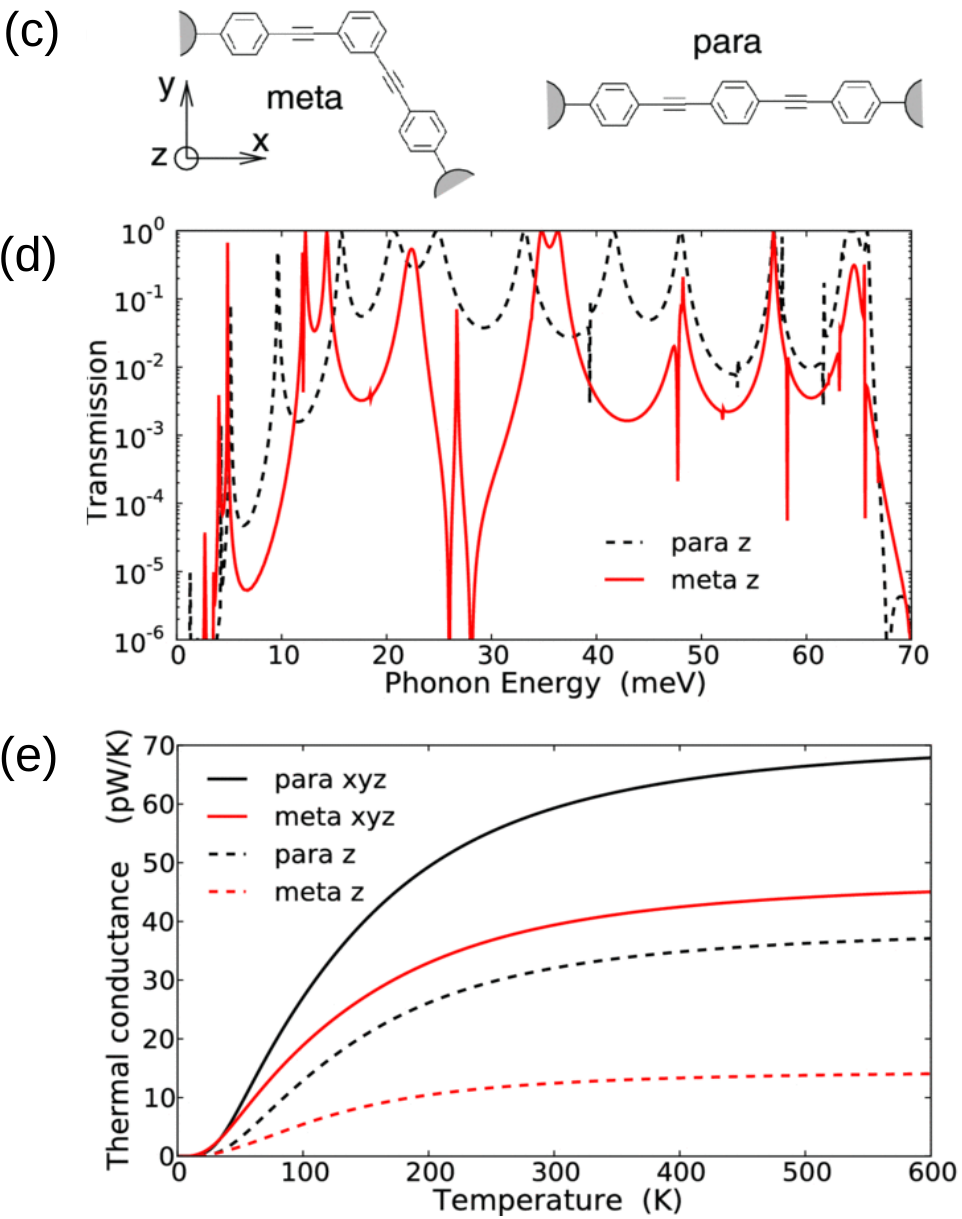}
  \end{minipage}
\caption{
Quantum interference effects of phononic conduction in molecular junctions.
(a)-(b) Examples of localized phonon modes (linear combinations of normal modes)
in oligo(phenylene-ethynlene) (OPE3), a cross conjugated molecule. Numbering counts the eigenmodes.
The modes may behave similarly to the Fano model (a) or the degenerate-levels model (b).
(c) meta and para OPE3 configurations, (d) calculated transmission functions, and
(e) the thermal conductance, considering out-of-plane modes only ($z$) or the complete spectrum involving ($x,y,z$) displacements.
Reprinted with permission from Reference \cite{markussen13},
Copyright (2013), AIP Publishing LLC. 
}
\label{figQI}
\end{figure}

\subsection{Applications}
\label{exampleH}

The transmission function (\ref{eq:Thar}) can be evaluated using
the so-called atomistic Green's function approach \cite{mingo}
to provide the thermal conductance of a broad range of nanoscale objects, 
carbon nanotubes, Si nanowires, graphene sheet,  graphite.
In such calculations one typically computes the phonon
dispersion based on a model for the interatomic potential, then evaluates the
transmission function employing a model for the self energies.
Phonon transport in carbon nanotubes was examined in e.g. Refs.  \cite{mingo05,CNT} using this analysis.

We discuss now examples of Landauer-based calculations of heat transfer in relatively short molecules.
Alkane chains and isotopically substituted disordered chains with 2-25 units
were examined in Ref. \cite{segal03}, showing the conductance to exhibit
non monotonic features at room temperature with growing size, see Fig. \ref{figAlkane}(a).
This behavior results from the fact
that the molecular vibrational spectrum may be largely altered with length for short molecules,
as well as the degree of localization of some molecular normal modes. 
Since details of
the molecule-heat baths (solids) coupling, the spectral properties of the solids, specifically the value of their 
Debye frequency, and the bulk temperature,
determine which molecular modes participate in the conduction process, 
the conductance of ultra-small systems can be tuned with size, before ballistic motion takes over  \cite{segal03}.
Fig \ref{figAlkane}(a) displays results at different temperatures
assuming ohmic thermal baths. 
Panel (b) depicts measurements of the thermal conductance in SAMs \cite{Gotsmann14}, 
demonstrating a qualitative agreement in trends
between theory and experiment.

Reducing phononic heat conductance
in molecular junctions (while maintaining high electronic conductance), is
desirable for improving thermoelectric efficiency.
Several design principles for molecules were tested in recent proposals, bringing modest success, as we discuss below.
The challenge in the manipulation of phononic current is rooted in the bosonic statistics:
Technically, if we replace the
bosonic form in Eq. (\ref{eq:landkappa})
by the corresponding fermionic function, the current would be determined by
the transmission function only in the vicinity of the equilibrium Fermi energy.
For molecular structures, the charge current then combines
contributions from a single or few resonance levels.
In contrast, the derivative of the Bose-Einstein distribution function in Eq. (\ref{eq:landkappa}) covers
frequencies in a window up to the baths' temperatures,
potentially comprising substantial
contributions from many molecular modes of different nature, localized and extended, to confuse dynamics.

In Ref. \cite{gemmaheat} it was proposed that molecules made of two separate subunits with a weak chemical link,
for example,  $\pi$-stacked aromatic rings, could show reduced thermal conductance relative to
the case with a single unit, while maintaining good electrical conductivity
(since electronic overlap through-space is preserved in $\pi-\pi$ systems).
Detailed simulations with first-principle parameters revealed a more intricate picture:
The phonon conductance may indeed reduce (by about a factor of 2) in this design,
relative to the case with a single unit, but in some cases, 
it {\it increased} due to an overall improvement of coupling in the junction.

Reduction of thermal conductance in molecules by employing the effect of quantum interference
was demonstrated in Ref. \cite{markussen13}. It was
pointed out that in e.g. a benzene ring with a meta connection to the leads,
the phononic transmission function (comprising in-plane vibrations) can exhibit destructive quantum
interference features, resulting in 
the suppression of the thermal conductance compared to the linearly conjugated analogue, see Fig. \ref{figQI}.
This behavior is conceptually similar to the electronic case \cite{Zant,Lambert15}. However,
while variations in electronic conductance due to interference effects may be of 1-3 orders of magnitude \cite{Zant,Lambert15},
changes in phononic conductance due to quantum interference are rather modest, factor of 1.5 to 5 
in  Ref. \cite{markussen13}. 
As explained above, this is because phononic conductance collects contributions from many modes of different nature, 
whereas quantum interference effects influence individually specific modes.

\section{METHODOLOGIES: Anharmonic Interactions}

The Landauer picture breaks down 
when deviations from the noninteracting normal-mode picture are significant.
This is the case in  systems with large-tunable anharmonicity [as in the FPU model \cite{FPU}] and at high temperatures
when inelastic phonon scatterings are influential.
What tools are available beyond the Landauer-harmonic approach?
Classical molecular dynamic simulations can be naturally applied beyond harmonic force fields, to include
interactions to all orders \cite{Lepri03,MDphonon,keblinski,galli}. 
Molecular dynamics simulations of heat conductance in simplified models
revealed the role of disorder and reduced dimensionality \cite{Dharhar,DharD1d,DharD2d} 
as well as anharmonicities \cite{leprian,DharFPU,Dhar3D,dharFPU14,bijay-an} on transport mechanisms and
the occurrence of nonlinear phenomena such as 
negative differential thermal conductance \cite{Li-NDTR}, thermal rectification \cite{Li-diode,Li-diodeCNT}, 
gating \cite{Li-transis}, and memory elements \cite{Li-memory}. 

Quantum mechanical treatments are typically limited to a certain range in parameters.
Among relevant methods 
we mention the non-equilibrium
Green's function technique, which is perturbative in the nonlinear interaction strength \cite{wang14,mingo}, and
master equation approaches, which may include the effect of anharmonicity exactly (as long as the model is minimal), but
are often limited to models with weak system-bath couplings \cite{juzarQME,SegalME06,SegalME09}.
Other schemes are
based on time-scale separation between slow and fast modes \cite{SegalBO},
phenomenological self-consistent treatments \cite{dhar08,roy06,Malay,Tulkki13},
and numerically-exact methodologies, computationally
limited to minimal modes \cite{MCTDHheatthoss,SaitoSB13,nazim}. 

\subsection{Self-Consistent Reservoirs}

\begin{figure} 
\includegraphics[scale=1]{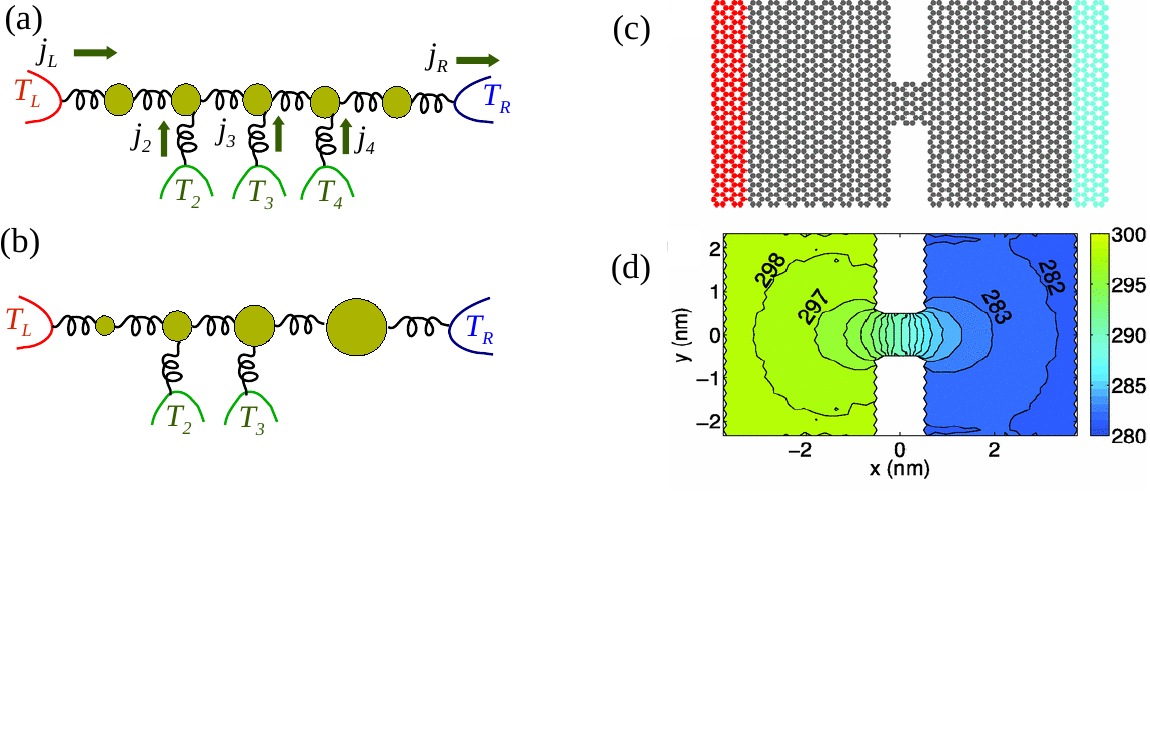}
\vspace{-20mm}
\caption{
(a) Illustration of the SCR model in a chain model with $N=5$ atoms.
Under the self consistent conditions, $j_L=j_R$. The conditions $j_2=j_3=j_4=0$ set the temperatures $T_{i}$, $i=2-4$.
(b) An example of a spatially asymmetric model (the different sizes reflect different masses), which realizes
a thermal diode effect - only in the quantum regime \cite{Malay}.
(c) Graphene nano constriction and
(d) its self-consistent temperature profile (K) in the quantum SCR model.
Temperatures were determined self-consistently for the gray atoms in the central region.
A classical approximation provided very similar results.
Panels (c)  and (d) are adapted with permission from Ref. \cite{Tulkki13}.
Copyrighted 2013 American Physical Society.
}
\label{figSCR}
\end{figure}

The self-consistent reservoir (SCR) method introduces effectively anharmonic interactions into harmonic models
by allowing the system to exchange energy with internal reservoirs which mimic many-body interactions.
In Fig. \ref{figSCR}(a)-(b) we exemplify the model with a one-dimensional chain.
The system includes $N$ atoms (not necessarily identical) in a junction setting,
connected to each other by harmonic springs. 
In addition, each particle is coupled to an independent harmonic (Langevin) thermal reservoir, also referred to as a
``thermal B\"uttiker probe" \cite{buttiker}.
The temperatures at the ``exterior" reservoirs, attached to the first and last particles, are set at $T_L$ and $T_R$, 
but the temperatures of the internal baths are determined in a self consistent manner,
by demanding that in steady state, on average, there is no net heat current from these baths to the physical system.

The SCR model has been investigated intensively with toy models.
It was originally proposed in the classical domain \cite{Bolsterli,Rich},
to prove the emergence of diffusional dynamics and Fourier's law of heat conduction upon the application of the
interior reservoirs  \cite{Bonetto04}.
The model was later extended to describe heat conduction in quantum systems, in linear response 
\cite{dhar08,RichQ,PereiraQSCR07}
and beyond that, analytically-approximately \cite{PereiraQCSCR11,PereiraPLA}, and numerically-exactly \cite{Malay,Tulkki13}.
What are the signatures of anharmonicity within the SCR model?
The interior reservoirs can gradually tune the transport dynamics, from ballistic-elastic to diffusive. Further,
the quantum SCR method can support a diode behavior (which is identically missing under the Landauer expression, as well as in the
classical SCR method), due to the interplay of quantum effects, spatial asymmetry, and effective anharmonicity, see
a possible setup in Fig. \ref{figSCR}(b) following Ref. \cite{Malay}.

Beyond toy models, the SCR method has been recently applied in Ref.  \cite{Tulkki13} for the study of quantum thermal 
transport in nanostructures such as the two-dimensional graphene constriction depicted in Fig. \ref{figSCR}(c).
The temperature profile in the central region was evaluated with a quantum SCR treatment, 
displayed in Fig. \ref{figSCR}(d).
Interestingly, it was found that 
the quantum profile closely matches classical results, suggesting that low-frequency modes,
for which quantum and classical statistics agree, largely contribute to transport in this system.

In conclusion, while missing genuine anharmonic interactions, the SCR method offers a simple mean
for studying phononic conduction in nanostructures while including both quantum effects and (effective) inelastic phonon scatterings.

\subsection{Non-equilibrium Green's Function Approach}
\label{NEGF}

\subsubsection{Formalism}

The non-equilibrium Green's function technique offers an elegant-powerful mean for computing equilibrium and out-of-equilibrium
properties of many-body systems \cite{schwinger, kadanoff, keldysh, haug}. 
For quantum transport problems, the NEGF approach provides a computational framework
for incorporating interactions (anharmonicities, in the heat transport problem), going beyond the Landauer expression.
We begin by introducing the Meir-Wingreen formula for the
steady state heat current flowing through an anharmonic system which is linearly
(harmonically) coupled to two harmonic reservoirs $L$ and $R$ \cite{wang14,MW, haug, MWthoss08},
\bea
j_q^L= \frac{1}{2\pi} \int_{0}^{\infty} d\omega\, \hbar \omega \, {\rm Tr}\big[G^{<}(\omega) \Sigma_L^{>}(\omega)-G^{>}(\omega) \Sigma_L^{<}(\omega)\big].
\label{eq:MW}
\eea
Here, $\Sigma_{\nu}^{<}(\omega)= -i\, n_{\nu}(\omega)\,\Gamma_{\nu}(\omega)$
and $\Sigma_{\nu}^{>}(\omega)= -i\, [1+n_{\nu}(\omega)] \,\Gamma_{\nu}(\omega)$,
are the self-energy components for $\nu=L,R$.
The harmonic part of the Hamiltonian was described in Sec. \ref{formalismH}. Adding nonlinear terms to $H_S$,
the molecular Green's function $G$ now incorporates anharmonicities. 
One can formally write down a closed set of equations for its components
in terms of the nonlinear self-energy $\Sigma_n$,
\bea
G^r&=& \big[(G_0^r)^{-1} -\Sigma_{n}^r\big]^{-1}= \big[(\omega+i0^{+})^2
\!-\!K_S \!-\!\Sigma_L^r\!-\!\Sigma_R^r\!-\!\Sigma_{n}^r\big]^{-1},
\nonumber\\
G^{<,>}&=& G^r\, \big(\Sigma_L^{<,>} +  \Sigma_R^{<,>} + \Sigma_{n}^{<,>}\big)\, G^a.
\label{eq:Keldysh}
\eea
For simplicity, we suppressed the $\omega$ dependence of these functions.
$G_0$ in Eq.~(\ref{eq:Keldysh}) precisely corresponds to the Green's function defined in the harmonic case
in Sec. \ref{theory-harmonic}.
The second equation in (\ref{eq:Keldysh}) is known as the ``Keldysh equation" \cite{keldysh}.
In the absence of anharmonic interaction, $\Sigma_n=0$, the Meir-Wingreen expression
reduces to the Landauer formula with the
transmission function Eq. (\ref{eq:Thar}).
In the interacting case it can be organized and expressed
in terms of an effective, temperature-dependent, transmission function \cite{wang-lu08,wang_meso},
\bea
\mathcal {T}_{\rm eff}(\omega, T_L, T_R)&\equiv&
{\rm Tr}\big[G^r \Gamma_L G^a \Gamma_R \big]\!
\nonumber\\
&+& \frac{1}{2(n_L\!-\!n_R)}{\rm Tr} \big[G^r \Gamma_n G^a (n_L \Gamma_L-n_R \Gamma_R) \!-\! i G^r \Sigma_n^< G^a(\Gamma_L-\Gamma_R)\big],
\nonumber \\
\label{eq:T-eff}
\eea
where $\Gamma_n=i (\Sigma_n^r-\Sigma_n^a)$ and $n_{\nu}$ denotes the Bose-Einstein distribution function of the
$\nu$ bath.
Under a symmetric coupling, $\Gamma_L=\Gamma_R=\Gamma$,
the transmission function reduces to
%
\bea
\mathcal{T}_{\rm eff}^{\rm sym}(\omega, T_L, T_R)= \frac{1}{2}{\rm Tr}\big[A(\omega)\Gamma(\omega)\big],
\eea
with $A(\omega)\equiv  i [G^r(\omega)-G^a(\omega)]$ as the spectral function of the molecule.

The Meir-Wingreen formula (\ref{eq:MW}) is formally exact,
but approximations should be employed to compute the nonlinear self-energy,
hence the current. Computational treatments include the
standard quantum field theory-type perturbative scheme
using the Feynman diagrammatic approach \cite{wang_meso,rammer_book, xu2008},  and
self-consistent approaches \cite{wang-lu08, mingo,luisier, wang_PRE, HOGalperin,mingo06}
in which the harmonic Green's function $G_0$ in $\Sigma_n$ is replaced by the full nonlinear $G$. 
In fact, the self-consistent approach under certain conditions provides results in agreement with
quantum master equation treatments \cite{juzar_12,zhang}, when the molecule-reservoir coupling is sufficiently weak.

What is the physical content of the retarded nonlinear self-energy $\Sigma_n^r$?
Its real part is responsible for frequency shifts of molecular vibrational
modes. The imaginary part corresponds to the finite life-time of phonons,
eventually responsible for the development of diffusive dynamics.
It should be mentioned though
that to fully realize the diffusive regime, high order phonon scattering processes should be included
in the nonlinear self-energy.
Once extracting the phonon lifetime $\tau_q$,
the thermal conductivity (denoted here by $\tilde \kappa$)
can be computed from the Boltzmann transport equation or more simply, using
the standard kinetic theory formula
$\tilde \kappa= \sum_{q} \frac{1}{3} c_q v_q^2 \tau_q$,  where $c_q$ is the heat capacity per unit volume of
mode $q$ and $v_q$ is the phonon group velocity \cite{mingo14ab}.


\begin{figure}
\includegraphics[width=0.7\columnwidth]{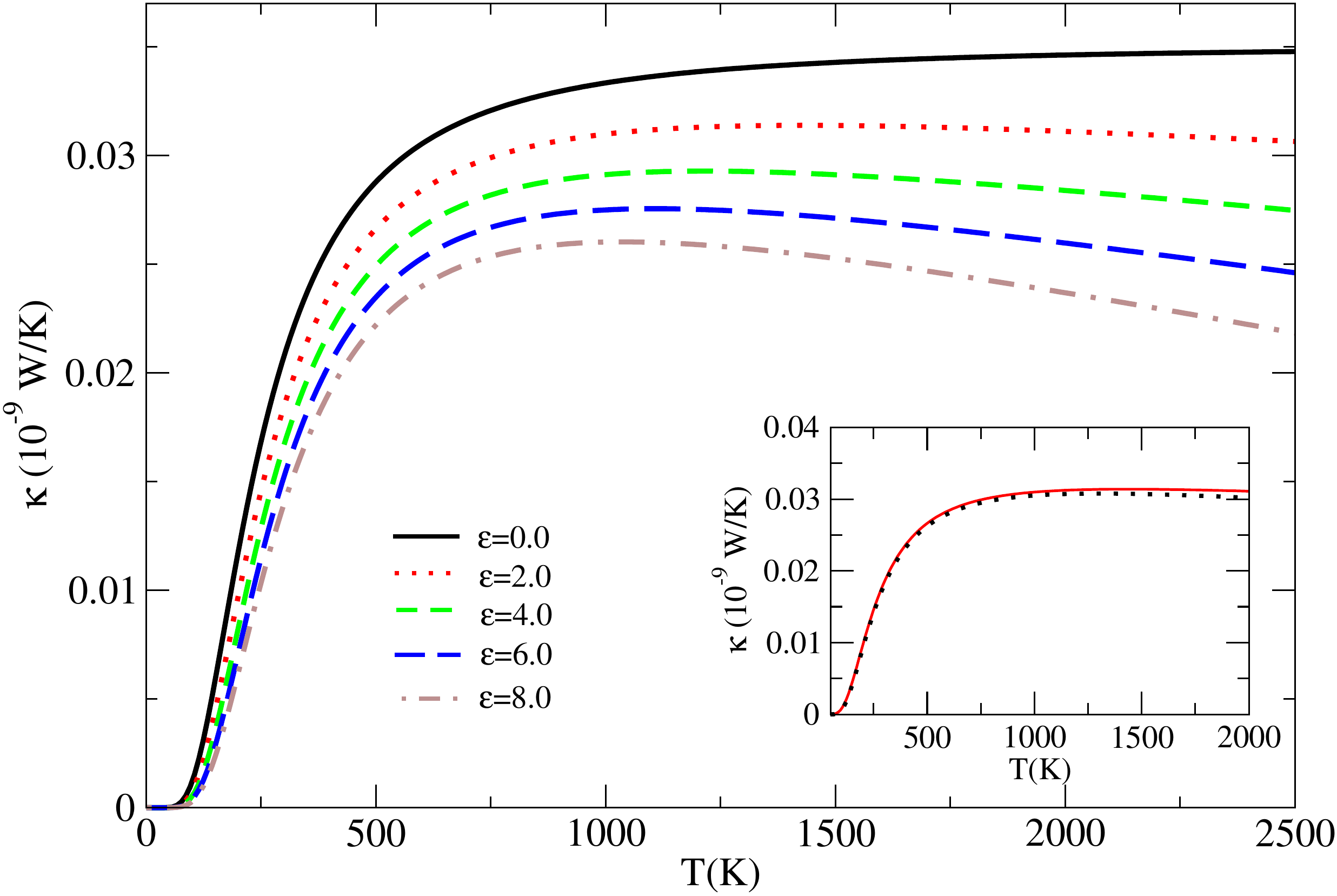}
\caption{Self-consistent NEGF calculations of the
thermal conductance as a function of temperature $T$ for an anharmonic junction.
The single harmonic oscillator junction ($\epsilon=0$) is supplemented by a quartic onsite potential
$H_n=\frac{\epsilon}{4}u_0^4$.
$\epsilon$ is given in units of
${\mathrm{eV/\left(amu^{2}\AA^{4}\right)}}$. 
The reservoirs are defined  by  semi-infinite chains of harmonic oscillators
with the nearest-neighbors force constant
$k=1\,\mathrm{eV/\left(amu\AA^{2}\right)}$ and an onsite element $k_{0}=0.1k$.
The ``molecule"  includes a single oscillator with
$K_{S}=1.1k$, and it is coupled to the left and right solids through
$\Lambda_{-1,0}^{LS}=\Lambda_{0,1}^{SR}=-0.25k$.
The inset compares the thermal conductance as calculated using the
self-consistent procedure (solid line) and the first-order expression (\ref{eq:T-quartic}) (dotted
line) for $\epsilon=2\,{\mathrm{eV/\left(amu^{2}\AA^{4}\right)}}$.
Adapted with permission from \cite{bijaythesis}.}
\label{fig_nonlinear}
\end{figure}

\subsubsection{Application}

The Landauer formula (\ref{eq:land}) was derived under the harmonic approximation
thus it rules out thermal rectification effects and other nontrivial nonlinear functionalities \cite{phononic}.
The NEGF formalism, in contrast, 
allows us to evaluate the role of anharmonicities in energy transport.
We exemplify the utility of the method on the single atom junction  [recall Eq. (\ref{eq:Tw0})],
allowing the oscillator to further interact with a quartic pinning potential
with the Hamiltonian $H_n= \frac{\epsilon}{4} u_0^4$ \cite{bijaythesis}.
To the lowest (first) order in $\epsilon$,
the components of the nonlinear self-energy are given by
$\Sigma_n^r(\omega) \!=\! \Sigma_n^a(\omega)\!=\!3 \,i \,\hbar \epsilon \, G_0^<(t=0)$,
$\Sigma_n^<(\omega)=\Sigma_n^>(\omega)=0$. Using these terms in Eq. (\ref{eq:T-eff}),
the transmission function with nonlinearities ${\mathcal T}_n(\omega)$
can be expressed in terms of the harmonic result ${\mathcal T}(\omega)$ 
\cite{huanan13},
\bea
{\cal T}_{n}(\omega;T_L,T_R) &=& \big(1+\Lambda(\omega;T_L,T_R)\big) {\cal T}(\omega), \nonumber \\
\Lambda(\omega;T_L,T_R) &=&  6i\, \hbar \, \epsilon \,G_0^{<}(t=0) \,{\rm Re}[G_0^r(\omega)].
\label{eq:T-quartic}
\eea
It should be recognized that the lesser component of the Green's function, $G_0^{<}(t=0)$, depends on temperature.
As a result, the effective transmission function becomes
temperature dependent and the junction can manifest nonlinear functionalities such as the diode effect. 

Fig. (\ref{fig_nonlinear}) displays the thermal conductance of this single anharmonic oscillator
junction, plotted against the averaged temperature $T$. Simulations were performed by employing
the self-consistent NEGF approach, replacing $G_0$ by the full $G$ in $\Sigma_n$, and solving
the Keldysh equation (\ref{eq:Keldysh}) iteratively, to compute the current from Eq. (\ref{eq:MW}) \cite{bijaythesis}.
It is evident that at low temperatures 
the junction's thermal conductance is dominated by elastic-harmonic forces.
In contrast, at high temperatures the thermal conductance is significantly reduced when the anharmonic pinning potential is turned on,
as inelastic phonon scattering processes largely influence transport.

The NEGF technique as described here is limited to small-atomistic models.
Efforts are dedicated to link and interface NEGF results
with coarse-gained modeling such as the Boltzmann transport equation.
This would allow simulations of larger nanoscale systems, and the resolution of e.g.
ballistic-diffusive crossover with parameter-free quantum mechanical first principle approaches \cite{mingo14ab}.


\subsection{Quantum Master Equation}

Projection Operator approaches, such as the Nakajima-Zwanzig projection operator technique  \cite{breuerbook},
trace out the dynamics of the (non-interesting) bath degrees of freedom, to reach equations of motion
for relevant-subsystem coordinates, typically, under the assumption of fast dynamics for the bath relative to the subsystem.
At the heart of such approaches is the conceptual separation of the model into three constituents: subsystem of interest, environment,
and an interaction term between the components.
Quantum master equations (QME) which describe the reduced dynamics of the density operator
are immensely useful for modelling
electron, proton, and exciton transfer, vibrational relaxation and spin dynamics in condensed phases \cite{nitzanbook}.
As well, in the related optical QME the radiation field plays the role of a thermal environment \cite{breuerbook}. 
Standard approximations involved in the framework of QME are: (i) weak subsystem-bath coupling, allowing
a perturbative expansion in the interaction strength.
(ii) Markov approximation, assuming short-lived memory effects in the bath, and (iii)
rotating wave approximation (often employed in the quantum optics literature),
leaving out off-resonant terms.

\begin{figure} 
\includegraphics[scale=0.4]{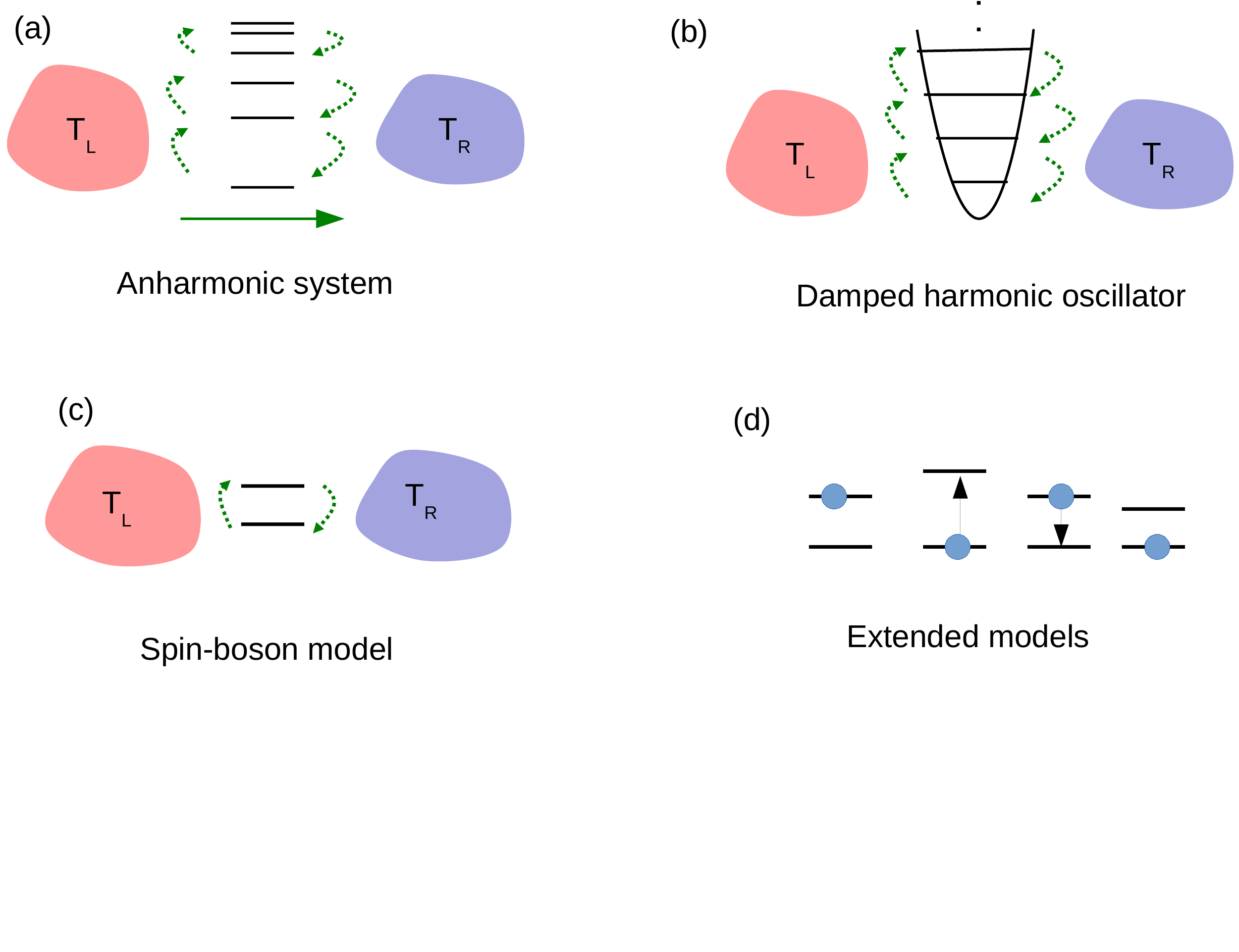}
\vspace{-15mm}
\caption{
Thermal nanojunctions in the energy representation for the molecular subsystem.
(a) Generic anharmonic junction.
(b) Damped harmonic oscillator model, generalized to the non-equilibrium case.
(c) Spin-boson model with an anharmonic, two-level system, coupled to harmonic baths.
(d) An extended-site model which can describe e.g. spin and exciton transfer in chains.
The arrows describe an energy transfer process, with a certain site being excited simultaneously with the
de-excitation of a different site 
}
\label{figHOTLS}
\end{figure}

QME traditionally communicate the dynamics, population and coherences, of a subsystem of interest.
More recently, efforts were put forward for
the development of related schemes for transport coefficients, to calculate
(charge, heat, spin) currents in non-equilibrium situations.
Advantages of QME over other approaches such as the NEGF are:
(i)  It can treat exactly intrinsic anharmonicities, yet obviously, limited to small ``impurity" models.
(ii)  Analytic results can be derived, to expose the role of different factors in the heat transport behavior.
The QME approach can be conveniently formulated in the energy basis of the subsystem. Some examples 
in the context of vibrational heat flow in molecular junctions
are illustrated in Fig. \ref{figHOTLS}.

\subsubsection{Weak molecule-reservoir coupling: The Bloch-Redfield-Markov equation}

We bring here a brief description of 
QME for vibrational heat transport in junctions,
under the assumption of weak molecule-bath coupling, by following Refs. \cite{SegalME06,SegalME09}.
Fig. \ref{figHOTLS}(a) displays the model, an $N$-state subsystem (possibly many-body nonlocal states) and two heat baths.
Excitation and relaxation processes in the subsystem are induced by complementing processes in the reservoirs.
The total Hamiltonian is 
\bea
H&=&H_S+ H_L+H_R+V_L+V_R,
\eea
where
\bea
H_S&=&\sum_{n}E_n|n\rangle \langle n|, \,\,\,\,  V_{\nu}=S^{\nu}\otimes B^{\nu}, \,\,\,\,\, \nu=L,R.
\eea
The energy spectrum of the subsystem (molecule) $E_n$ may be derived from a truly anharmonic potential without further approximations.
In the damped harmonic oscillator model of Fig. \ref{figHOTLS}(b), $E_n=n\hbar\omega_0$.
The molecular system is coupled to thermal baths $H_{\nu}$, $\nu=L,R$,  not necessarily harmonic [see examples in
Refs. \cite{Segalradiation,SegalME09}],
with the subsystem operators $S^{\nu}=\sum_{m,n}S^{\nu}_{m,n}|m\rangle \langle n|$ and 
the operator $B^{\nu}$, given in terms of the $\nu$ bath degrees of freedom.
Leaving out details, it can be shown that the population $P_n$ of the subsystem eigenstates obey kinetic equations \cite{SegalME09}
\bea
\dot P_n(t)=-P_n(t) \sum_{\nu,m} |S^{\nu}_{n,m}|^2k_{n\rightarrow m}^{\nu}
+ \sum_{\nu,m}|S_{n,m}^{\nu}|^2 k_{m\rightarrow n}^{\nu} P_m(t),
\label{eq:PtimeQME}
\eea
with rate constants
\bea
k_{n\rightarrow m}^{\nu}=\int_{-\infty}^{\infty} e^{-iE_{n,m}\tau/\hbar}\langle B_{\nu}(\tau)B_{\nu}(0)\rangle_{T_{\nu}}d\tau.
\label{eq:rateQME}
\eea
This result is correct to the lowest nontrivial order in the subsystem-bath coupling. It was derived under a Markovian approximation,
and after decoupling the diagonal and off-diagonal elements of the reduced density matrix.
The operators in Eq. (\ref{eq:rateQME}) are given in the interaction representation, $\langle... \rangle_{T}$
represents an average with respect to the equilibrium distribution of the two baths,  $E_{m,n}\equiv E_m-E_n$.
Applying similar steps, the steady state heat current across the system [depicted in Fig. \ref{figHOTLS}(a)]
explicitly follows \cite{SegalME06,SegalME09}
\bea
j_q=\frac{1}{2}\sum_{n>m}E_{m,n} |S_{m,n}^{L}|^2 [P_n k_{n\rightarrow m}^{L}-P_mk_{m\rightarrow n}^{L}
]
- \frac{1}{2}\sum_{n>m}E_{m,n} |S_{m,n}^{R}|^2 [P_n k_{n\rightarrow m}^{R}-P_mk_{m\rightarrow n}^{R}
].
\nonumber\\
\label{eq:jQME}
\eea
Here, the population $P_n$ are the long-time solution of Eq. (\ref{eq:PtimeQME}).
This symmetric expression is quite intuitive: The current is given by calculating the net transfer process from all transitions
between the $n$ and $m$ states. Rate constants are multiplied by energy differences and 
weighted by the population of states.
A more general result, involving contributions from coherences, was derived in Ref. \cite{juzarQME}.

We exemplify this expression in the ``atomic limit", with a junction comprising of a single oscillator.
In the harmonic limit, using the bosonic creation $b_{\nu,q}^{\dagger}$ and annihilation $b_{\nu,q}$ operators,
$H_{\nu}=\sum \hbar \omega_q b_{\nu,q}^{\dagger}b_{\nu,q}$, $H_{S}=\hbar \omega_0b_0^{\dagger}b_0$,
$V_{\nu}=\sum_{\nu,q}\frac{\hbar \tilde \lambda_{\nu,q}}{2}(b_{\nu,q}^{\dagger}+b_{\nu,q})(b_{0}^{\dagger}+b_0)$, one can derive
the Landauer-like result, 
\bea
j_q= \hbar\omega_0\frac{\gamma_L(\omega_0) \gamma_R(\omega_0)}{\gamma_L(\omega_0)+\gamma_R(\omega_0)} \left[n_L(\omega_0)-n_R(\omega_0)\right],
\label{eq:jHO}
\eea
where $\gamma_{\nu}(\omega)=\pi/2 \sum_{q}\tilde\lambda_{\nu,q}^2 \delta(\omega-\omega_q)$ stands for the mode-bath coupling energy,
consistent with the definitions of the damping function in Sec. \ref{formalismH}. 
Eq. (\ref{eq:jHO}) can be obtained from Eq. (\ref{eq:jqsingle}) taking the limit $\gamma_{\nu}\ll \omega_0$,
and it describes a resonant-sequential transmission process, as only reservoirs' modes which match 
the molecular frequency $\omega_0$ scatter through the junction.

Beyond the harmonic approximation, the ``spin boson model"
is a simple yet truly nontrivial example for anharmonic heat conduction. In this case,
the subsystem spectrum is truncated to include only two states with a gap $\omega_0$, see Fig. \ref{figHOTLS}(c).
The spin-boson Hamiltonian includes a two-state ``impurity",
 $H_S={\hbar\omega_0\over 2} \sigma_z$, interacting with harmonic baths
$H_{\nu}=\sum \hbar \omega_q b_{\nu,q}^{\dagger}b_{\nu,q}$ according to
$V_{\nu}=\sigma_x\sum_{q} {\hbar\tilde \lambda_{\nu,q}\over 2}
 (b_{\nu,q}^{\dagger} +b_{\nu,q})$.
Here $\sigma_x$ and $\sigma_z$ are the Pauli matrices.
Using Eqs. (\ref{eq:PtimeQME})-(\ref{eq:jQME}), one receives now the heat current \cite{segal05SB,SegalME06}
\bea
j_q= \hbar\omega_0
\frac{\gamma_L(\omega_0) \gamma_R(\omega_0)}{\gamma_L(\omega_0)[1+2n_L(\omega_0)]
+\gamma_R(\omega_0)[1+2n_R(\omega_0)]} \left[n_L(\omega_0)-n_R(\omega_0)\right].
\label{eq:jTLS}
\eea
%
Since the baths temperatures enter the transmission probability here,
this junction can display the thermal diode effect when spatial asymmetries are presented \cite{segal05SB}.
This nonlinear functionality is a definite fingerprint of anharmonic terms.

The weak-coupling second order QME presented here can be extended systematically to higher orders, to account e.g. for
co-tunneling processes \cite{juzar4}.
Insight can be also earned by writing down Fermi-golden rule rates for multi-phonon scattering processes, then constructing
the overall heat transfer rate, or the thermal current, based on the framework of kinetic master equations 
\cite{Leitner,Leitner99,straubbook}.
For a cubic anharmonicity, for example, Fermi golden rule rates consist two types of (energy conserving) processes:
a vibrational mode may decay into two others, or two modes can collide to produce a third mode. This type of analysis
is particularly useful for simulating vibrational energy transfer in amorphous materials, glasses \cite{leitnerglass}, 
proteins \cite{straubbook,leitnerProtein03,leitnerProtein05,straub} and across interfaces \cite{leitner-interface}. 

\subsubsection{Strong molecule-surface coupling: polaronic equations}

Equation (\ref{eq:jQME}) was derived under the assumption of a
molecule weakly attached to the thermal reservoirs. It predicts
a linear enhancement of the thermal conductance with increasing coupling energy $\gamma$, see for example Eq. (\ref{eq:jHO}).
Obviously, at strong coupling, corresponding to a molecule tightly attached to the interface, this trend should break down.
By using the so-called non-interacting-blip approximation (NIBA) \cite{weissbook},
which is related to a polaronic treatment \cite{Dekker},
it was recently demonstrated that the thermal conductance of the anharmonic spin-boson model (with an ohmic dissipation)
displays a crossover behavior with the system-bath coupling parameter, satisfying
$\kappa \propto \frac{k_B\omega_0^2}{\omega_c}\left(\frac{\hbar \omega_c}{k_BT}\right)^{1-2\alpha}$;
$\alpha=\alpha_L+\alpha_R$ \cite{SegalPRE14}. 
We used the ohmic damping function,
$\gamma_{\nu}(\omega)=\pi\alpha_{\nu}\omega e^{-\omega/\omega_c}$
with $\omega_c$ the Debye (cutoff) frequency of the phononic baths and $\alpha$ a dimensionless parameter.
It is also useful to define 
$E_r\equiv2\alpha_{\nu}\omega_c$ as the reorganization-interaction energy of the molecular mode with the baths' phonons.
This NIBA result for the conductance agrees with numerically exact Monte-Carlo simulations, at high temperatures \cite{SaitoSB13}.

\begin{figure} 
\hspace{9mm}
\includegraphics[scale=0.28]{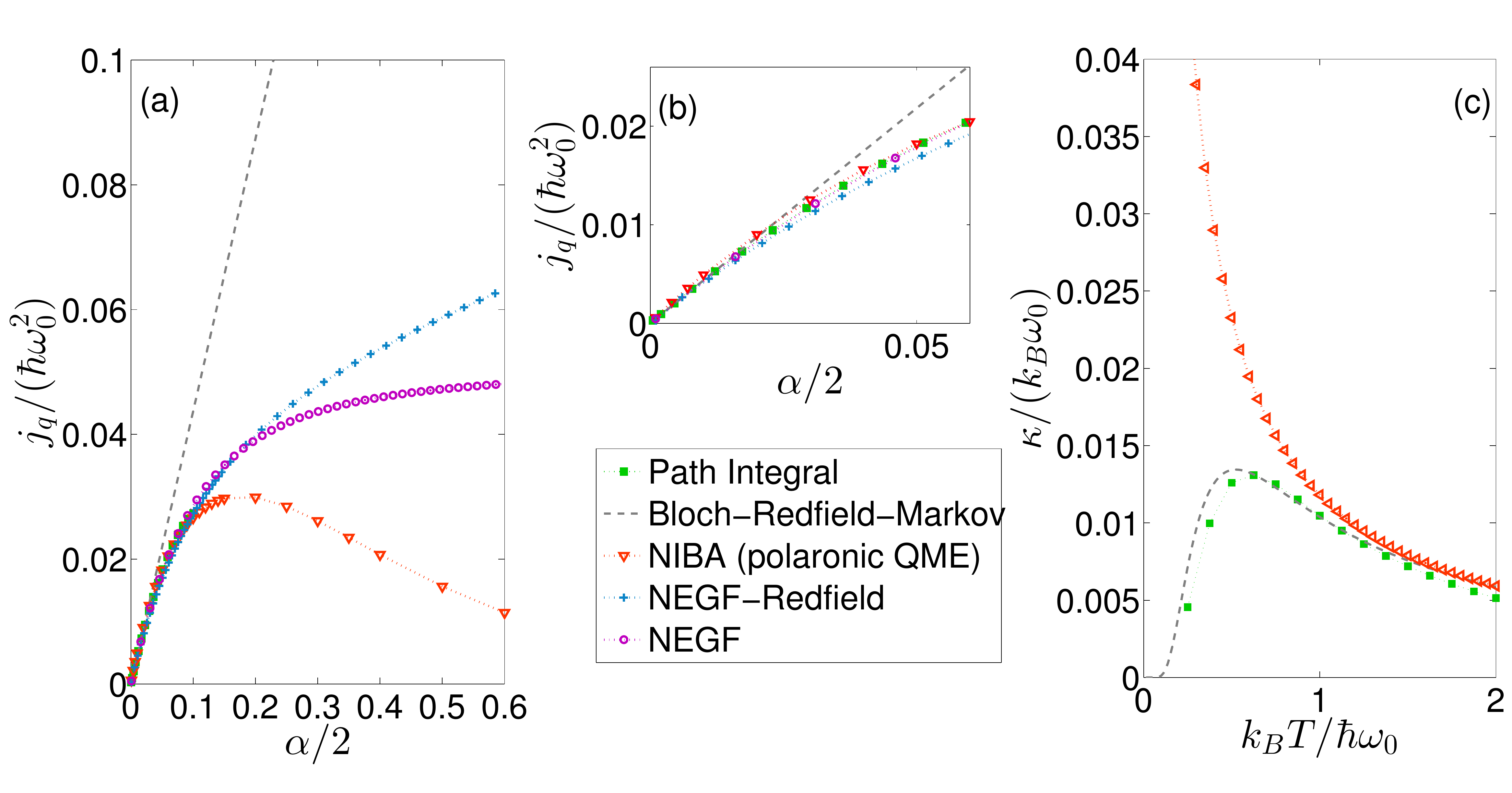}
\caption{Heat transfer in an anharmonic nanojunction, the spin-boson model of Fig. \ref{figHOTLS}(c).
(a) Current as a function of the molecule-bath coupling parameter $\alpha=\alpha_L+\alpha_R$,
demonstrating the breakdown of the linear
scaling predicted by the weak coupling Bloch-Redfield-Markov QME \cite{nazim}, with a suppression of the current at strong coupling.
Panel (b) Zooms over the weak coupling regime where different approximations agree.
$k_BT_L=2\omega_0$ and  $k_BT_R=\omega_0$, $\omega_c/\omega_0=10$.
(c) Thermal conductance $\kappa=j_q/\Delta T$ as a function of temperature in the linear response regime, $\Delta T=T_L-T_R$.
NIBA (polaronic QME) correctly behaves only at high temperatures.
$\omega_c/\omega_0=10$ and $\alpha_{L,R}=0.0172$. For more details, see Ref. \cite{nazim}.
}
\label{figQMEstrong}
\end{figure}

Fig. \ref{figQMEstrong} displays the heat current in the spin-boson model, a representative of an anharmonic junction,
 under different approximation schemes:
a standard weak-coupling treatment (Bloch-Redfield-Markov QME),
NIBA-polaronic QME, an NEGF-Redfield approach \cite{MWthoss08} and a more accurate variant of NEGF \cite{NEGFwu14},
as well as numerically-exact influence functional path integral simulations \cite{segalINFPI10},
performed based on the mapping of the model onto a fermionic description. For more details, see Ref. \cite{nazim}.
We make the following observations:
(i) The trend $j_q \propto \alpha$ fails beyond the very-weak coupling limit.
(ii) At high temperatures, $k_BT>\hbar \omega_0$, the thermal conductance
decays with the increase of temperature, 
a manifestation of the intrinsic anharmonicity of the junction.
This trend is correctly captured by QME calculations, see Fig. \ref{figQMEstrong}(c).
In contrast, at low temperatures this trend is
reversed since thermal occupation factors of bath modes
dominate the behavior, rather than temperature-dependent
(anharmonic) effects on the junction. NIBA simulations, however, fail to capture this turnover behavior.
Approximate techniques which can bridge between different regimes, to consistently describe transport e.g. at weak and strong coupling,
are of great interest \cite{QME-NIBA}, as well as the derivation of bounds for heat transport in anharmonic junctions \cite{Ed}.

\subsection{Numerically-Exact Simulations}

The methods detailed so far 
are limited to describe heat conduction in certain physical regimes due to their underlying approximations,
as exemplified in Fig. \ref{figQMEstrong}.
It is highly desirable thus to develop first-principle techniques which can go beyond perturbation theories, to
reveal the crossover between different regimes and to provide benchmarks for approximate theories.
Obviously, such nonperturbative treatments are numerical in nature and computationally limited to the simulation 
of minimal nanojunctions such as the two-bath spin-boson model. 
The dissipative dynamics of a two-state system coupled to a single heat bath, the (original) spin-boson model,
has been examined in many studies  \cite{Leggett}. Among the numerically exact machinery employed to treat it we list
the quasi-adiabatic propagator path-integral (QUAPI) approach \cite{QUAPI1,QUAPI2},
which builds upon the Feynman-Vernon influence functional, the multiconfiguration time-dependent Hartree
(MCTDH)  method, a wavefunction theory \cite{thossMCTDH}, and Monte Carlo (MC) simulations \cite{MC-SB}.
These tools were developed to hand over the dynamics of the reduced density matrix of the subsystem.
However, a current operator is a more complex object in nature since it is explicitly defined in terms 
of the baths degrees of freedom.

Generalizations of numerically exact tools 
to simulate phononic heat current were reported in several recent studies:
The multilayer multiconfiguration time-dependent Hartree theory was extended in Ref. \cite{MCTDHheatthoss}, and it revealed a
turnover behavior of the current with increasing coupling strength, similarly to Fig. \ref{figQMEstrong}(a).
Influence-functional path integral simulations were employed in Refs. \cite{nazim,segal13qubit},
by mapping boson environments into fermion baths \cite{segalINFPI10}.
Monte Carlo simulations of the (linear response) thermal conductance were performed in \cite{SaitoSB13}, by
mapping the spin-boson model onto an Ising model with long-range exchange interactions. 
More recently, influence function path integral expressions were generalized to describe heat exchange between a
subsystem and a heat bath  \cite{Weissheat}, though a calculation of the steady state heat current under a
QUAPI-type approach is still missing.

These numerically exact treatments are limited to a minimal anharmonic heat conducting nanojunction,
a qubit, yet they are immensely useful for our understanding of quantum transport in hybrid junctions \cite{Nori} 
in a broad range of parameters:
under weak/strong coupling of the subsystem to the heat baths, low/high temperatures, slow/fast subsystem dynamics
relative to the baths' motion,  and when different baths are employed,
e.g., with rich spectral properties, possibly taking into account bath anharmonicities.
Future studies will aim to bridge the gap between minimal models and more physical realizations of anharmonic junctions.

\section{Conclusions}
In this review, we focused on the problem of vibrational energy transfer (or thermal/heat/phononic conduction)
in a junction geometry with a molecule bridging two heat reservoirs. 
We described recent experiments on vibrational heat transfer in molecular junctions focusing on alkane chains. We then
reviewed methodologies useful for calculating thermal conduction in the quantum domain,
starting from coherent-harmonic phononic conduction in nanojunctions, and concluding with highly-anharmonic yet minimal
qubit constructions.
We began by recounting the Landauer approach, appropriate for the study of elastic transport.
This simple tool can be readily employed to
simulate phononic conduction in extended systems, as long as the molecular normals modes
and their hybridization with the contacts are given.
The Landauer-elastic scattering picture breaks down when interactions (anharmonicities) are presented.
We outlined several techniques which account for anharmonic effects:
The self-consistent reservoirs method mimics inelastic scattering of phonons with temperature probes.
The non-equilibrium Green's function technique as described here accounts for genuine many-body effects, albeit perturbatively, and 
it is limited to simple-minimal models.
Reduced density matrix projection operator techniques can be extended to
study thermal conduction of ``impurity models",  when a molecule is strongly coupled to macroscopic solids.
Finally, numerically exact techniques can simulate heat exchange
in minimal models such as the spin-boson nanojunction, 
providing benchmarks for approximate techniques.

Developing techniques for computing quantum heat transfer in nanostructures,
bridging the gap between minimal-model junctions and real systems, is highly desirable.
Ongoing efforts are dedicated towards the development of 
semiclassical techniques in which quantum statistics is implemented within classical molecular dynamics simulations
\cite{wang07QC,semi15},
construction of quantum corrections to classical expressions \cite{MDphonon,Amon}, and
computations using Boltzmann transport equation with input from first-principle atomistic approaches \cite{mingo14ab}.

In this review we only considered conduction of heat due to vibrations of nuclear degrees of freedom.
However, when a molecule links metal electrodes,
electrons as well considerably participate in the heat conduction process.
Understanding how the interaction between electrons and atomic vibrations affect electron dynamics as well as
energy transfer in real systems is a fundamental-formidable task, 
groundwork for making further progress in molecular nanotechnology.

%


\section*{ACKNOWLEDGMENTS}
This work was funded by the Natural Sciences and Engineering
Research Council of Canada, the Canada Research Chair Program
and the CQIQC at the University of Toronto. 
We thank Renante Yson for assistance in generating graphics.

%

\bibliography{heatrev.bib}

\bibliographystyle{ar-style3.bst}

\end{document}